# Quantum oscillations of magnetoresistance of the submicrometer thick bismuth telluride-based films

L.N. Lukyanova[*], Yu.A. Boikov, V.A. Danilov, O.A. Usov, M.P. Volkov, V.A. Kutasov

Ioffe Physical-Technical Institute, Russian Academy of Sciences, Polytekhnicheskaya 26, 194021 St.-Petersburg, Russia

*Corresponding author: e-mail lidia.lukyanova@mail.ioffe.ru, Phone: +7 812 5159247, Fax: +7 8125154767

Hetero-epitaxial films based on bismuth telluride with excess of Te were grown by hat wall technique at the surface of the mica (muscovite). Galvanomagnetic properties of the thin films were measured, and quantum oscillations of the magnetoresistance were found at the temperatures below 10 K in the magnetic field from 6 to 14 T.
From analysis of magnetoresistance oscillations the main surface state parameters of the films were determined.

Expementally obtained Landau level index shift and its temperature dependence are consistent with Berry phase specific for topological Dirac surface states. The estimated parameters of electronic topological surface states of the bismuth telluride-based films are of special interest because of possible usage of them in micro generators and micro coolers, and also for other device applications.

## Introduction

Topological insulators, predicted theoretically [1] and observed experimentally in $Bi_2Te_3$ [2, 3] and its alloys [4-6], is a new quantum state of matter that shows insulator state in the bulk and conducting states at the surface, the spin of surface states electron is being locked perpendicular to its momentum due to strong spin-orbit coupling. As the result, the surface state electrons demonstrate dissipation-less charge transport with high mobility because they cannot back scattering on defects [7, 8]. But in practice the bulk is not a perfect insulator, for instance antisite defects or vacancies introduced during the process of films growth.

To minimize the bulk transport contribution, submicron $Bi_2Te_3$ films were grown with Te excess using hat wall technique. These films were used for measurements and analysis of the magnetoresistance oscillations, that permit to obtain the cyclotron resonance frequency, cross-section Fermi surface, Landau level indexes, effective cyclotron mass, Fermi wave vector, velocity and Fermi energy, surface charge carrier concentration, lifetime and mobility of charge carriers.

## 1 Oscillations of Magnetoresistance of $Bi_2Te_3$+Te films

Hetero-epitaxial films $B_2Te_3$+Te were grown by hat wall technique at the surface of the mica (muscovite) [9].

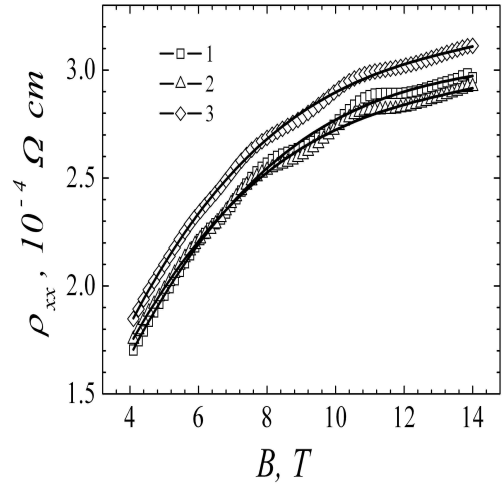

Figure 1. Experimental magnetoresistance dependence $\rho_{xx}$ (1-3) and smoothed background lines (trend) of the $\rho_{xx}$ on magnetic field B. Figure 1 plots $\rho_{xx}$ at the temperatures: 1 – 1.6 K, 2 – 4.2 K, 3 – 10 K.

The films consist of the single crystal blocks which are well c-axis preferentially oriented. The transport properties of the films were measured at low temperatures.

The experimental dependences of magnetoresistivity $\rho_{xx}$ of $Bi_2Te_3$ +Te film with the 0.8 μm thickness for temperatures at 1.6 K, 4.2 K and 10 K on magnetic field $B$ are shown in Fig. 1, curves 1-3.

The magnetoresistance quantum oscillations were clearly observed with magnetic fields more than 6 T for all three temperatures. The curves 1-3 in Fig. 2 show magnetoresistance oscillations $\Delta\rho_{xx}$ on inverse magnetic field $1/B$ after subtracting smoothed background.

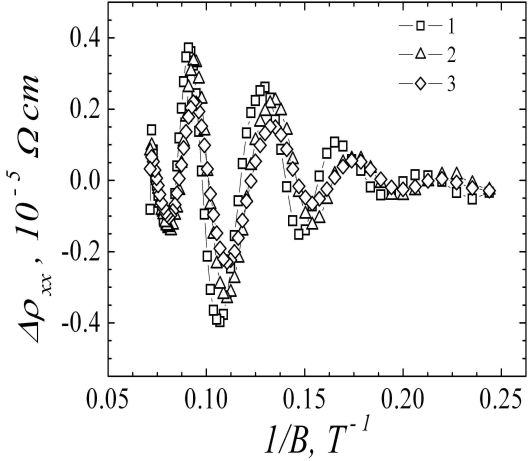

Figure 2. Magnetoresistance quantum oscillations $\Delta\rho_{xx}$ (1-3) dependence on inverse magnetic field $1/B$ after subtracting background.

Figure 2 plots $\Delta\rho_{xx}$ at the temperatures: 1 - 1.6 K, 2 – 4.2 K, 3 – 10 K.

## 2. Cyclotron resonance frequency of the magnetoresistance quantum oscillations

The frequency dependence of normalized amplitude of cyclotron resonance lines were obtained by fast Fourier transformation (FFT) method using the experimental quantum oscillations of magnetoresistance shown in Fig. 2, curve 1-3.

Figure 3 revealed only one resonance frequency for each selected temperatures. The amplitudes of the cyclotron resonance decrease with increasing temperatures from 1.6 K to 10 K, they being equal to 23, 24 and 25 T for temperatures 10, 4.2 and 1.6 K, respectively.

The difference between calculated cyclotron resonance frequencies (Fig. 3) lies within an experimental errors so for cyclotron resonance frequency it was accepted its mean value $F_{cr}$= 24 T.

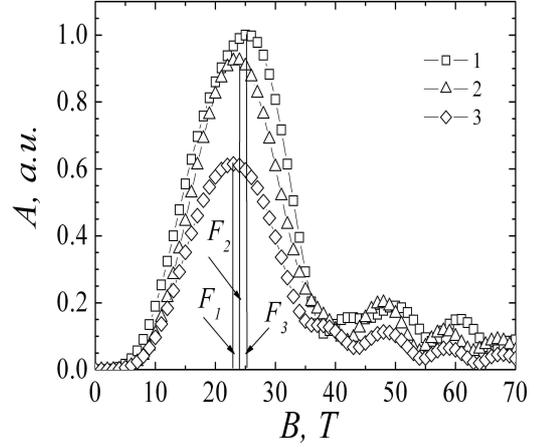

Figure 3. Normalized amplitude of the cyclotron resonance $A$ (1-3) dependence on frequency $F$ obtained by FFT.
Calculated cyclotron resonance frequency $F_{cr}=F(A=Amax)$ (vertical lines ): $F_1$, – 25T at 1.6 K, $F_2$ – 24T at 4.2 K, $F_3$ – 23 T at – 10 K, respectively.

## 3. Fermi sphere parameters

In accordance with the Onsager relation [3], cyclotron resonance frequency $F$ is proportional to the cross section of the Fermi surface $S(k_F)$ that is given as [10]:

$$F = \left(\frac{\hbar}{2\pi e}\right) S(k_F) \qquad (1)$$

where, $k_F$ is the Fermi wave vector. For spherical Fermi surface the wave vector $k_F$ is obtained as

$$k_F = \sqrt{\frac{S(k_F)}{\pi}} \qquad (2)$$

The surface concentrations of charge carriers $n_s$ was determined using next relations:

$$n_s = 4\pi k_F^{\,2} \qquad (3)$$

The Table 1 contains the frequency $F$ of cyclotron resonance, the cross-section of Fermi surface $S_F$, the wave vector $k_F$, the surface concentrations of charge carriers $n_s$. The data given in the Table 1 are in agreement with the results of the Refs. [3, 4].

Table 1. Fermi surface parameters and charge carrier surface concentration $n_s$ in the $Bi_2Te_3$ +Te films

| $F$, T | $S(k_F)$, $nm^{-2}$ | $k_F$, nm | $n_{fs}$ $10^{12}$, $cm^{-2}$ |
|---|---|---|---|
| 24 | 0.23 | 0.27 | 0.58 |
| 30 [4] | 0.29 | 0.30 | 0.72 |
| 41.7 [3] | 0.40 | 0.36 | 1.0 |
| 50 [4] | 0.48 | 0.39 | 1.21 |

### 4. Landau levels

The magnetoresistance oscillations in high magnetic fields and low temperature occur due to the redistribution of carriers among the Landau levels when one of the Landau levels, separated by the cyclotron energy, crosses the Fermi energy.

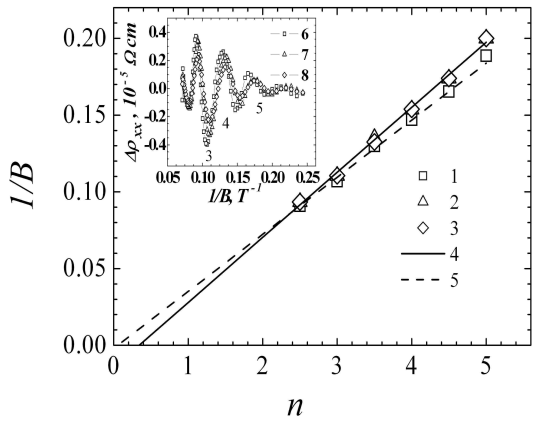

Figure 4. Dependences of Landau level indexes $n$ (1-3) on magnetic field and corresponding them linear extrapolated lines (4, 5) that determine index shift $\gamma$. Insert shows the dependence of magnetoresistance oscillations $\Delta\rho_{xx}$ (6-8) on inverse magnetic field $1/B$ and oscillation minima in $\Delta\rho_{xx}$ with corresponding Landau level indexes at temperatures T: 1.6 K - 1, 6, 4.2K - 2, 7, 10 K - 3, 8.

The period of the oscillations is given by [10]:

$$2\pi(n+\gamma) = S(k_F) \cdot \frac{\hbar}{eB} \quad (4)$$

The phase shift of Landau level index was evaluated using linear extrapolation of experimental Landau level index dependence on inverse magnetic field in limit at $1/B$=0, Fig. 4.

The experimental values of the phase shift of the Landau level index depend on temperatures: the phase shift $\gamma$ =0.1 at 1.6 K, and the phase shift $\gamma$ =0.38 at 4.2 K and 10 K. The phase shift $\gamma$ small deviates from exactly $\pi$-Berry phase, as also reported on other studies of topological Dirac surface states [3, 12-14].
The deviation of the observed from the value specific for $\pi$-Berry phase has been variously attributed to the Zeeman effect and the distortion of linear dispersion of the Dirac cone related to the surface states [13, 14].

### 5. Cyclotron effective mass

The cyclotron effective mass $m^*_{cyc}$ were obtained by least square fitting of the normalized experimental temperature dependence of maximal amplitude of the magnetoresistance oscillations $\Delta\rho_n(T)$ and calculated function $\Delta\rho_c(T, m^*_{cyc})$, the $\Delta\rho_n(T)$ and $\Delta\rho_c(T, m^*_{cyc})$ being determined as:

$$\Delta\rho_n(T) = \frac{\Delta\rho_{xx}(T_{n=1-3})}{\Delta\rho_{xx}(T_1)} \quad (5),$$

$$\Delta\rho_c(m^*_{cyc}, T, B) = \frac{\sinh(X)}{X}, \quad (6)$$

where $X(m^*_{cyc}, T, B) = c_{LK} m^*T/B$, effective cyclotron mass $m^*_{cyc}=m_{cyc}/m_0$, $m_0$ – electron mass; constant $c_{LK}=2\pi^2 m_0/(\hbar e)$; $\hbar$ = reduced Plank constant.

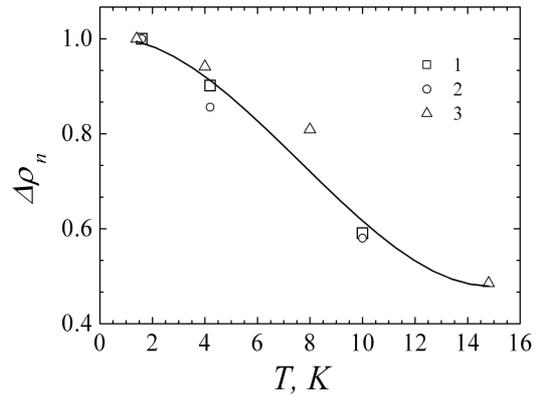

Figure 5. The temperature dependence of the maxima of the normalized amplitude of magnetoresistance oscillations $\Delta\rho_n$.
1 – 1-st maximum, 2 – 2-nd maximum, 3 – [4].

The normalized amplitude of oscillations Δ(n(T) selected for each experimental temperature curve Tn=1-3 at T1 = 1.6 K, T2 = 4.2 K, T3 = 10 K) as shown in Fig. 5.

The cyclotron effective mass equal $m^*_{cyc}$ = 0.3, estimated with equations (5-6) from analysis of Fig. 5, is found to be in a good agreement with other published data [4].

### 6. Charge carrier relaxation lifetime, mean free path and mobility

The relaxation lifetime τ of charge carrier was obtained by least square linear fitting of the Dingle function $Y_{calc}(\tau, 1/B, T)$ and experimental data $Y_{exp}(1/B, T)$ dependence on the inverse magnetic field $1/B$.

Dingle function $f_D$ is given as [5, 10]:

$$f_D(\tau) = e^{\frac{-\pi}{\omega_{cyc}\tau}} \quad (7)$$

where

$$\omega_{cyc} = \frac{e}{m_{cyc}} \cdot \frac{1}{B} \quad (8)$$

Experimentally based function $Y_{exp}(1/B, T)$ is determined as:

$$Y_{exp}\left(\frac{1}{B}, T\right) = \ln\left(\frac{\Delta\rho_{xx}}{\rho_0} \cdot \frac{\sinh(X)}{X}\right), \quad (9)$$

where $\Delta\rho_{xx}$ is the magnetoresistance oscillations in the magnetic field $1/B$=0.1, corresponding to the maximal amplitude magnetoresistance oscillations at the Landau level index $n$=5; $\rho_0$ is the resistivity; $\sinh(X)/X$ is the Lifshits-Kosevich function.

After obtaining lifetime, the values of Dingle temperature $T_D$, mean free path $l_F$ and mobility $\mu$ of charge carriers (Table 2) were determined using relations:

$$T_D = \frac{h}{k_b} \cdot \frac{1}{\tau} \quad (10)$$

$$l_F = v_F \tau \quad (11)$$

$$\mu = \frac{e\tau}{m_{cyc}} \quad (12)$$

where $k_b$ is Boltzman constant; $v_F$ is Fermi velocity.

Table 2. Cyclotron resonance frequency $v= \omega_{cyc}/2\pi$, Fermi wave vector $k_{ef}$, Fermi velocity $v_F$ and Fermi energy $E_F$, charge carrier mobility $\mu$, mean free path $l_F$, and Dingle temperature $T_D$.

| $v$, THz | $k_F$, nm$^{-1}$ | $v_F$, 10$^5$, m/s | $E_F$, meV |
|---|---|---|---|
| 2.0 | 0.276 | 1.06 | 19.3 |
| $\mu$, m$^2$/Vs | $l_F$, nm | $\tau$, 10$^{-13}$, s | $T_D$, K |
| 0.33 | 59 | 5.5 | 2.2 |

The obtained parameters shown in the Table 2 are special interest because permit to evaluate the contribution of topological Dirac surface states to transport properties of the layered thermoelectrics based on Bi$_2$Te$_3$.

### Conclusions

In conclusion, we studied galvanomagnetic properties of hetero-epitaxial Te-excess bismuth telluride films grown by hat wall technique. Quantum oscillations of the magnetoresistance were detected at low temperatures T below 10 K in the range of magnetic field from 6 to 14 T.

As evidence from Landau diagram, the experimentally obtained Landau level index shift is temperature dependent and changed in the range 0.1 - 0.4, which is consistent with a nontrivial Berry phase and topological Dirac surface states. From analysis of the magnetoresistance quantum oscillations the surface state parameters such as cyclotron resonance frequency, charge carrier mobility, charge carrier surface concentration, Fermi velocity and energy, and mean free path were determined.

The unique topological Dirac surface states of the investigated films make them perspective for innovations including thermoelectric micro generators and coolers, and multifunctional heterostructures for device applications.

### Acknowledgments

This study is partially supported by the Russian Foundation for Basic Research, project no. 13-08-00307a. Staff of the International Laboratory of low temperatures and high magnetic fields (Wroclaw, Poland) is acknowledged for technical assistance during measurements of the film magnetotransport parameters.

## References

[1]. C.L. Kane, E.J. Mele. Phys. Rev. Lett. **95**, 246802 (2005)
[2]. Y. L. Chen, J.-H. Chu, J. G. Analytis, Z. K. Liu, K. Igarashi,4 H.-H. Kuo, X. L. Qi, S. K. Mo, R. G. Moore,1 D. H. Lu,1 M. Hashimoto, T. Sasagawa, S. C. Zhang, I. R. Fisher, Z. Hussain, Z. X. Shen, Science **329**, 359 (2010).
[3]. Dong-Xia Qu, Y. S. Hor, Jun Xiong, R. J. Cava, N. P. Ong. Science **329**, 821 (2010).
[4]. A. Taskin, Z. Ren, S. Sasaki, K. Segawa, and Y Ando. Phys. Rev. Lett., **107**, 016801 (2011).
[5]. P. Gehring, B. F. Gao, M. Burghard, and K. Kern. NanoLett. **12**, 5137 (2012)
[6]. Liang He, Faxian Xiu, Xinxin Yu, Marcus Teague, Wanjun, Jiang, Yabin Fan, Xufeng Kou, Murong Lang, Yong Wang, Guan Huang, Nai-Chang Yeh, and Kang L. Wang, Nano Lett. 12, 1486 (2012)
[7]. M.Z. Hasan, C.L. Kane, Rev. Mod. Phys., 82, 3045-3067 (2010).
[8]. J.H. Bardarson, and J.E Moore, Rep. Prog. Phys. **76,** 056501 (2013)
[9]. 9 Yu.A. Boikov, V.A. Danilov. Tech. Phys., **53**, 3, 348, (2008).
[10]. D. Shoenberg. Magnetic Oscillations in Metals. Series: Monographs on Physics. Cambridge University Press, Cambridge (2009) 596 p.
[11]. Z. Ren, A. Taskin, S. Sasaki, K. Segawa, Y Ando. Phys. Rev. B, **82**, 241306 (2010).
[12]. J. G. Analytis, R. D. McDonald, S. C. Riggs, J.-H. Chu, G. S. Boebinger, I. R. Fisher. Nat. Phys., **6**, 12, 960 (2010).
[13]. A.R. Wright, R. H. McKenzie. Phys. Rev. B **87**, 085411/11 (2013).
[14]. J.N. Fuchs, F. Pirechon, M.O. Goerbig, and G. Montambaux. Eur. Phys. J. B **77**, 351 (2010).